\documentclass[12pt, letterpaper]{article}
\usepackage[top=1in, bottom=1.5in, left=1in, right=1in]{geometry}
\usepackage[utf8]{inputenc}
\usepackage{booktabs}
\usepackage{natbib}
\usepackage{hyperref}
\usepackage{graphicx}
\usepackage{authblk}

\title{A Cadence to Reduce Aliasing in LSST}

\author[ ]{Keaton J.\ Bell$^1$, Kelly M.\ Hambleton$^2$, Michael B.\ Lund$^3$,  \\ and R{\'o}bert Szab{\'o}$^{4,5}$, \\ with the support of the LSST Transient and Variable Stars Collaboration} 
\affil[1]{Max Planck Institute for Solar System Research}
\affil[2]{Villanova University}
\affil[3]{Vanderbilt University}
\affil[4]{MTA CSFK, Konkoly Observatory, Hungary}
\affil[5]{MTA CSFK Lend\"ulet Near-Field Cosmology Research Group}

\date{November 2018}

\begin{document}

\maketitle

\begin{abstract}
Regular sampling in the time domain results in aliasing in the frequency domain that complicates the accurate determination of the periods of astrophysical variables.  We propose to actively break the regularity of this sampling by providing an additional consideration for the scheduler that weights fields according to when observations will contribute the least to aliasing. The current aliases for each field can be computed during daytime from the history of observations. We can then calculate the times when additional observations would worsen or alleviate these aliases for different fields. The scheduler should give preference to observation epochs that lessen the effect of aliasing, while still meeting all other cadence requirements.
\end{abstract}

\section{White Paper Information}
Correspondence can be directed to Keaton Bell: \href{mailto:bell@mps.mpg.de}{bell@mps.mpg.de} 
\begin{enumerate} 
\item {\bf Science Category:} Exploring the Transient Optical Sky

\item {\bf Survey Type Category:} the main `Wide-Fast-Deep'
   survey
\item {\bf Observing Strategy Category:} An integrated program with science that hinges on the combination of pointing and a detailed observing strategy. 
\end{enumerate}

\clearpage

\section{Scientific Motivation}

LSST will obtain a thorough record of the variable sky.  The most fundamental time domain quantity of physical interest is the timescale of photometric variations.  Accurate periods of cyclic processes are essential for many areas of astrophysics, e.g., probing the interiors of stars that undergo global pulsations, or revealing the physical properties of components in stellar binaries.  LSST's usefulness for constraining the properties of new variable sources will be limited by its data's ability to deliver accurate period determinations.

The considerations that determine the recoverability of intrinsic periods from irregular, sparse survey photometry are quite different than for classical, continuous time series data.  In particular, gaps in the observations cause samples in Fourier power spectra to be strongly correlated.  By introducing uncertainty in the number of cycles missed during the gaps, multiple candidate frequencies can compete to describe the data equally well, with only one corresponding to the intrinsic frequency of physical interest. These difficulties are conflated by noisy measurements, where beating of these aliases against noise can cause incorrect peaks to appear dominant. The more structured the time sampling is, the more the aliases can confuse the period determination.  We propose that the LSST scheduler should be mindful of aliasing and work to distribute its sparse observations with minimal regularity.

\begin{figure}[h]
  \centering
  \includegraphics[width=0.8\columnwidth]{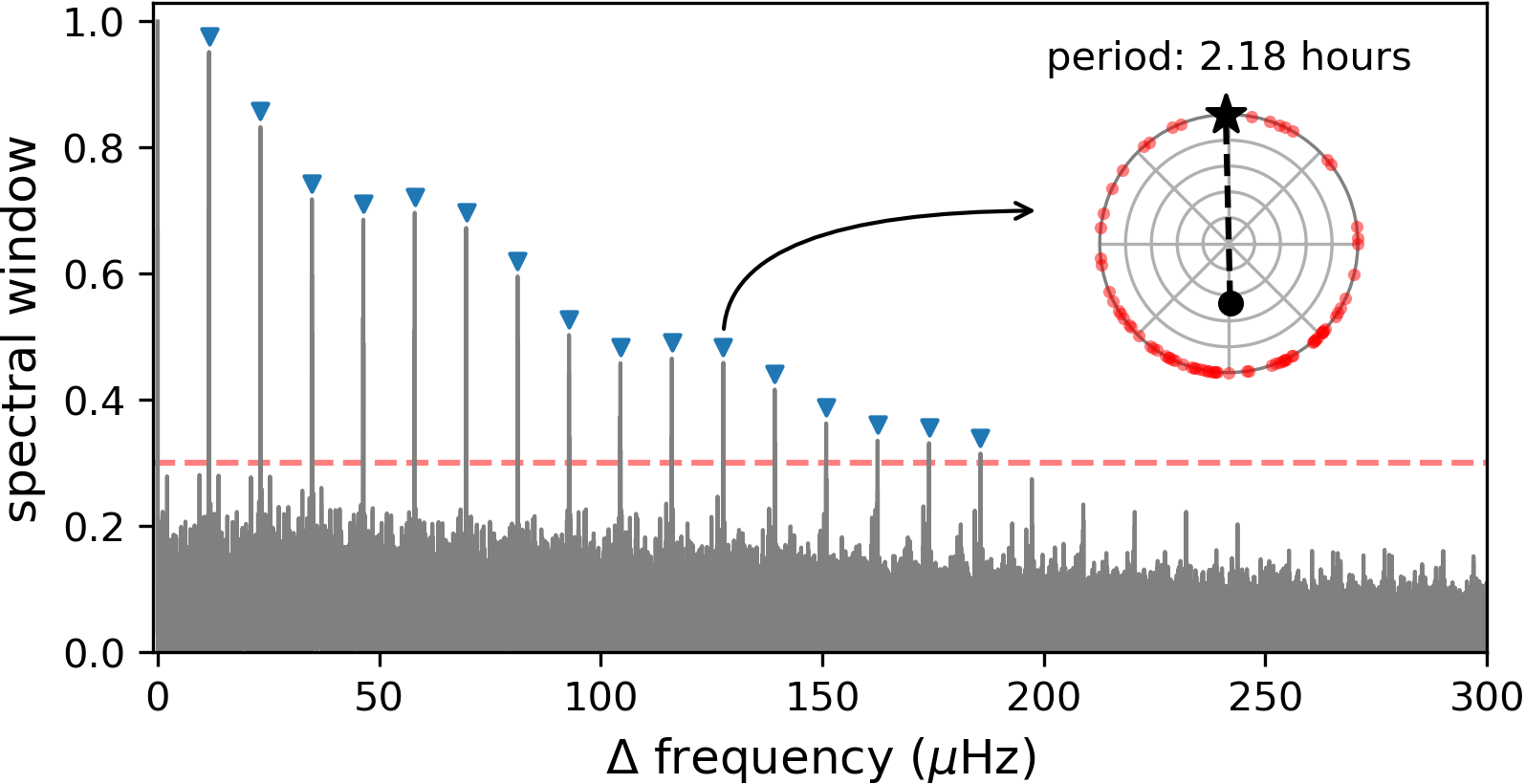}
  \caption{Spectral window after 3 years of observations for an example pointing of the Wide-Fast-Deep survey in the minion\_1016 simulated baseline cadence. The inset plot shows the concentration in phase of observations that produce one of the alias peaks \citep[following][]{Eyer1999}. The star symbol indicates the phase of a future observation that would most reduce the existing alias. Figure~\ref{fig:changes} shows the effect of future observations on all of the marked aliases.}
  \label{fig:lsstsw}
 
\end{figure}

Figure~\ref{fig:lsstsw} displays the spectral window---the characteristic signature about a signal in a Fourier transform given the time sampling---produced by three years of simulated observations from the baseline cadence, OpSim \citep{Delgado2014} simulation minion\_1016 \citep[ignoring filter information;][]{LSST2017}.  The significant pattern of aliases is dominated by the diurnal peaks at $\approx11.6\,\mu$Hz caused by observations being acquired only during nighttime.  But not all of these peaks correspond to the unavoidable diurnal constraints; the highlighted alias peak is caused by a concentration in the phase of observations on a timescale of 2.18 hours. \citet{Eyer1999} introduced this intuitive interpretation that the amplitude of an alias peak is equivalent to the offset of the center-of-mass of all time samples when they are folded on the alias period and placed on a unit circle (as in the polar plot in Figure~\ref{fig:lsstsw}).  The star marker in anti-phase of this center-of-mass marks the timing of a future observation that would best reduce the amplitude of this particular alias.

The overall LSST Wide-Fast-Deep cadence is defined to achieve multiple overarching science goals.  For instance, fields are ideally revisited on timescales between 15 and 60 minutes to measure the motion of Solar System objects \citep{LSST2009}.  This is a fairly strict survey constraint, but by admitting a range of acceptable revisit times that accomplish the intended science, it is also flexible enough to allow excessive structure in the observations to be avoided.  We propose a modification to the main LSST Wide-Fast-Deep survey cadence whereby the scheduler considers the existing aliases in each survey field and times new visits within the acceptable science-driven windows to actively reduce their amplitudes.  This will not change the general timescales being sampled as motivated by LSST's science goals, but will aid the interpretation of the data for meeting these goals.  Beyond reducing alias amplitudes, this will also result is better sensitivity to detecting variability at what would otherwise be alias frequencies with poor phase coverage.

\begin{figure}
  \centering
  \includegraphics[width=1\columnwidth]{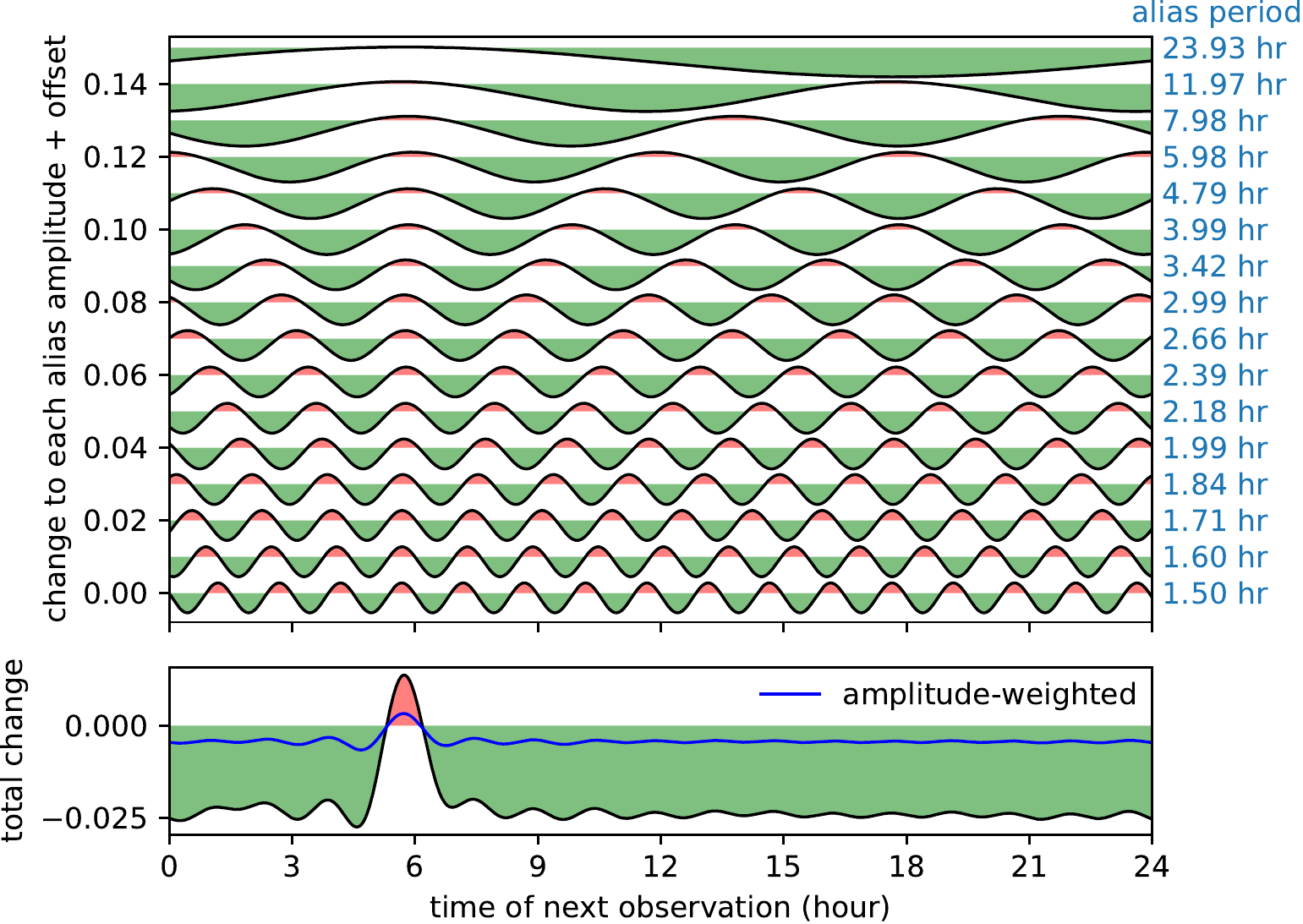}
  \caption{The effect of applying this method to a single field using the example depicted in Figure~\ref{fig:lsstsw}. Top: The effect on each individual alias, where green depicts a reduction in aliasing and red depicts an increase. Bottom: the total change in alias amplitudes where the blue line is the amplitude-weighted total. It can be clearly seen that an observation at the 6 hr mark would be detrimental to the spectral window. New observations made outside of the existing $\sim1$-hour nightly concentration of revisits will result in a net reduction of aliasing overall.}
  
  \label{fig:changes}
\end{figure}

The top panel of Figure~\ref{fig:changes} demonstrates the amount by which amplitudes of each of the largest aliases marked in the example from Figure~\ref{fig:lsstsw} will change in response to a single new observation over a 24-hour period.  The existing concentration in phase of observations that produces the observed set of strong aliases is visible where all of these curves are simultaneously positive near 5.8 hours.  The bottom panel shows the net change in the alias amplitudes as a straight sum (green and red filled), and as a sum weighted by the current alias amplitudes (blue).  The aliasing is exacerbated overall if new observations reinforce the existing phase concentration within the $\sim1$-hour nightly window near 5.8 hours, but they are alleviated overall for any new observations outside this small window.  We propose that this blue curve in the bottom panel of Figure~\ref{fig:changes} be used as a factor that the scheduler considers when deciding the timing of visits for each field.  This will limit the severity of aliasing for time domain astronomy with LSST and will actively reduce the amplitudes of any aliases that develop.

The Lund et al.\ white paper also discusses the value of minimizing any potential aliasing on a more cursory level as part of broader science goals.

\clearpage

\vspace{.6in}

\section{Technical Description}

\subsection{High-level description}

The LSST scheduler algorithm must account for many factors when deciding which field is to be observed next.  The time since the last observation is one of the most important, as LSST seeks to monitor the variable sky on a variety of timescales.  The scheduler will also presumably take into account airmass, seeing, filter history, etc.\ in deciding its next field to target.  We propose that among this set of considerations for selecting future fields, the present set of aliases be accounted for. Field revisits should be given preference through the weighting scheme when new observations would reduce aliasing, downweighting fields when new observations would worsen the aliasing.  

The current aliasing in a field can be evaluated by taking the Fourier transform of a sequence of constant values sampled at the specific observation history.  The Fourier transform should be sampled $\approx10$ times more densely than the frequency resolution (inverse of the total observational baseline) to get an accurate reading of the alias amplitudes.  Each of the aliases that exceed some significance threshold should be identified, and the amount that future observations would collectively reduce these alias amplitudes (offsets of all of the alias points in their polar phase plots, as in Figure~\ref{fig:lsstsw}) should be included as a weight in the scheduler.  The influence of new observing times on the full set of significant aliases can be computed by the sum of adjustments to each, weighted by their current amplitudes.  This is demonstrated for simulated observations of an example field by the blue line in the bottom panel of Figure~\ref{fig:changes}. By accounting for the existing aliases (which can be computed for each field during the daytime), the timing of all future visits are informed by the entire past history of observations, rather than only the time since the other most recent visits.  This additional consideration for the scheduler can be combined with any other Wide-Fast-Deep cadence proposals to maximize the science yield from the data.

\subsection{Footprint -- pointings, regions and/or constraints}
This scheme should be used when executing the full Wide-Fast-Deep survey, and potentially also minisurveys.  It will not significantly affect the survey footprint or total number of observations, but will simply stagger the timings of the field visits to avoid too much regularity in revisit times that will hamper the measurements of variability timescales.

\subsection{Image quality}

N/a.

\subsection{Individual image depth and/or sky brightness}

N/a.

\subsection{Co-added image depth and/or total number of visits}

N/a.

\subsection{Number of visits within a night}

N/a. 

\subsection{Distribution of visits over time}
The distribution of visits over time should avoid as much structure as possible while still sampling the various timescales of interest to meet the science objectives.

\subsection{Filter choice}

N/a.

\subsection{Exposure constraints}

N/a.

\subsection{Other constraints}

N/a.

\subsection{Estimated time requirement}

Same time requirements as whatever baseline Wide-Fast-Deep survey is adopted that this scheme makes adjustments to.

\begin{table}[ht]
    \centering
    \begin{tabular}{l|l|l|l}
        \toprule
        Properties & Importance \hspace{.3in} \\
        \midrule
        Image quality & 3    \\
        Sky brightness & 3 \\
        Individual image depth & 3  \\
        Co-added image depth & 3  \\
        Number of exposures in a visit   & 3  \\
        Number of visits (in a night)  & 2  \\ 
        Total number of visits & 3  \\
        Time between visits (in a night) & 1 \\
        Time between visits (between nights)  & 1  \\
        Long-term gaps between visits & 1 \\
        Other (please add other constraints as needed) & 3 \\
        \bottomrule
    \end{tabular}
    \caption{{\bf Constraint Rankings:} Summary of the relative importance of various survey strategy constraints. Ranked from 1=very important, 2=somewhat important, 3=not important.}
        \label{tab:obs_constraints}
\end{table}

\subsection{Technical trades}
To aid in attempts to combine this proposed survey modification with others, the following questions are addressed:
\begin{enumerate}
    {\it \item What is the effect of a trade-off between your requested survey footprint (area) and requested co-added depth or number of visits?}
    
    The survey footprint will not be affected by the proposed modification.
    {\it \item If not requesting a specific timing of visits, what is the effect of a trade-off between the uniformity of observations and the frequency of observations in time? e.g. a `rolling cadence' increases the frequency of visits during a short time period at the cost of fewer visits the rest of the time, making the overall sampling less uniform.}
    
    The average time between revisits should not be affected, but the revisits should be less uniformly spaced.  This modification should be explored as an addition to various other survey strategies under consideration, including a rolling cadence.
    {\it \item What is the effect of a trade-off on the exposure time and number of visits (e.g. increasing the individual image depth but decreasing the overall number of visits)?}
    
    A greater number of observations can achieve better phase coverage of more timescales and is preferred.
    {\it \item What is the effect of a trade-off between uniformity in number of visits and co-added depth? Is there any benefit to real-time exposure time optimization to obtain nearly constant single-visit limiting depth?}
    
    This proposal is concerned only with adjusting the timings of observations in a way that improves their interpretability for time domain science.  That said, real-time adjustments to the exposure times would alter the amount that variations during each measurement are smoothed, which could introduce a different complication to interpreting light curves of variable sources.
    {\it \item Are there any other potential trade-offs to consider when attempting to balance this proposal with others which may have similar but slightly different requests?}
    
     Evaluating the amount of aliasing in simulated surveys will give the final word on how difficult specific strategies will make the interpretation of the time domain data. The only trade-offs would be to cadences that strive for strict regularity, which we argue would inhibit many areas of time domain astronomy.
\end{enumerate}

\section{Performance Evaluation}

The severity of aliasing in simulated LSST data can be evaluated from survey strategy realizations simulated with OpSim \citep{Delgado2014} using the Metrics Analysis Framework \citep[MAF;][]{Jones2014}.  The proposed strategy modification is expected to be useful by the authors, but this needs to be tested by its implementation into the scheduler in simulated OpSim runs.  While we expect this approach to decrease the severity of aliasing, it is not obvious that these observation-by-observation adjustments represent the optimal scheme for minimizing aliases, and we encourage other strategies to also be proposed and simulated. \citet{Lund2016} introduced a basic MAF metric to measure the strength of aliases in simulated data, but this needs to be modified to account for the full set of severe aliases produced by the survey (see Figure~\ref{fig:lsstsw}), and to ensure that the computed spectral window is sufficiently oversampled in frequency so as to detect all significant aliases.  The median of a value such as the sum of the amplitudes of the highest 20 alias peaks in the spectral window across every field would be a useful summary statistic to try to minimize.

\section{Special Data Processing}

No special data processing is required, though additional computations based on timestamps from the standard processing pipeline would need to be considered by the scheduler.

\section{Acknowledgements}

This work was developed within the Transients and Variable Stars Science Collaboration (TVS) and the authors acknowledge the support of TVS in the preparation of this paper.

\section{References}

\begin{itemize}
\bibitem[Delgado et al.(2014)]{Delgado2014} Delgado, F., Saha, A., Chandrasekharan, S., et al.\ 2014, Proceedings of the SPIE, 9150, 915015 
\bibitem[Eyer \& Bartholdi(1999)]{Eyer1999} Eyer, L., \& Bartholdi, P.\ 1999, A\&AS, 135, 1 
\bibitem[Jones et al.(2014)]{Jones2014} Jones, R.~L., Yoachim, P., Chandrasekharan, S., et al.\ 2014, Proceedings of the SPIE, 9149, 91490B 
\bibitem[LSST Science Collaboration et al.(2009)]{LSST2009} LSST Science Collaboration, et al.\ 2009, arXiv:0912.0201 
\bibitem[LSST Science Collaboration et al.(2017)]{LSST2017} LSST Science Collaboration, et al.\ 2017, arXiv:1708.04058 
\bibitem[Lund et al.(2016)]{Lund2016} Lund, M.~B., Siverd, R.~J., Pepper, J.~A., \& Stassun, K.~G.\ 2016, PASP, 128, 025002 
\end{itemize}

\end{document}